\documentclass[aps,prb,twocolumn,superscriptaddress,showpacs,amsfonts]{revtex4-1}
\usepackage{epsfig}
\usepackage{graphicx,graphics}
\usepackage{amsmath,ulem}
\usepackage{amssymb}
\usepackage{color}
\usepackage{bm}
\begin{document}

\title{Curvature-induced Rashba spin-orbit interaction in strain-driven nanostructures}
\author{Paola Gentile}
\affiliation{CNR-SPIN, I-84084 Fisciano (Salerno), Italy and
Dipartimento di Fisica ``E. R. Caianiello'', Universit\`a di
Salerno, I-84084 Fisciano (Salerno), Italy}
\author{Mario Cuoco}
\affiliation{CNR-SPIN, I-84084 Fisciano (Salerno), Italy and
Dipartimento di Fisica ``E. R. Caianiello'', Universit\`a di
Salerno, I-84084 Fisciano (Salerno), Italy}
\author{Carmine Ortix}
\affiliation{Institute for Theoretical Solid State Physics, IFW
Dresden, D-01069 Dresden, Germany}

\begin{abstract}
We derive the effective dimensionally reduced Schr\"odinger
equation with spin-orbit interaction in low-dimensional electronic
strain driven nanostructures. A method of adiabatic separation
among fast normal quantum degrees of freedom and slow tangential
quantum degrees of freedom is used to show the emergence of a
strain-induced Rashba-like spin-orbit interaction (SOI). By
applying this analysis to one-dimensional curved quantum wires we
demonstrate that the curvature-induced Rashba SOI leads to
enhanced spin-orbit effects.
\end{abstract}


\maketitle

\section{Introduction}
The interest in the theory of quantum physics on bent manifolds
has received a boost since the experimental progresses in
synthesizing low-dimensional nanostructures with curved geometries
-- the next generation nanodevices \cite{ahn06,ko10,mei07,par10}.
From a purely theoretical point of view, the problem of the
quantum motion of a particle living in a curved space has
represented a matter of controversy for a long time
\cite{dew57,jen71,dac82,kap97}. The problem arises because Dirac
quantization on a curved manifold leads to operator-ordering
ambiguities\cite{dew57}. However, Jensen and Koppe \cite{jen71}
and later da Costa \cite{dac82} (JKC) have introduced a thin-wall
quantisation procedure that circumvents this pitfall. It treats
the quantum motion on a curved two-dimensional (2D) surface -- or
analogously on a torsionless planar one-dimensional (1D) curved
manifold -- as the limiting case of a particle in
three-dimensional (3D) space subject to lateral quantum
confinement. For both 1D curved manifold and 2D surface with a
translational invariant direction, the JKC method allows to
eliminate the surface curvature from the effective dimensionally
reduced Schr\"odinger equation at the expense of adding a
potential term to it. The problem is therefore substantially
simplified, since quantum carriers can actually live in a 2D or 1D
space in the presence of a curvature-induced quantum geometric
potential (QGP). On the nanoscale, the QGP can actually cause many
intriguing phenomena
\cite{can00,aok01,enc03,fuj05,kos05,gra05,cha04,mar05}. For
instance, a quantum particle constrained to a periodically curved
surface senses a periodic QGP acting as a topological
crystal\cite{aok01}. Likewise the QGP in spirally rolled-up
nanostructures leads to winding-generated bound
states\cite{ved00,ort10}. The experimental observation of these
fascinating phenomena is hindered by the fact that in actual
systems with curvature radii on the order of a few hundred
nanometers, the QGP is still very weak -- it is in magnitude
proportional to $\hbar$ --  and typically only comes into play on
the sub-Kelvin energy scale. It has been recently shown
\cite{ort11b} that by accounting for the effect of the deformation
potentials of the model-solid theory \cite{van89,sun10}, the
strain field induced by the curvature leads to a strain-induced
geometric potential (SGP) which has the same functional form as
the QGP but strongly boosting it. One cannot therefore distinguish
between curvature-induced quantum effects and curvature-induced
classical strain effects in nanosystems -- they are always present
and contribute in different amount to the same geometric
potential.

In this work, we show that this strain field of bent
nanostructures leads to a curvature-induced Rashba spin-orbit
interaction whose strength follows precisely the local curvature
of the manifold. This is  immediately relevant for electronic
nanodevices since the modern nanostructuring
method\cite{ahn06,ko10} is based on the tendency of thin films
detached from their substrates to assume a shape yielding the
lowest possible elastic energy. As a result, thin films can either
roll up into tubes\cite{sch01,pri00} or undergo wrinkling to form
nanocorrugated structures\cite{fed06,mei07,cen09}. The nanoscale
variation of the strain represents a key property of such bent
nanostructures. It leads to considerable band-edge
shifts\cite{den10} with regions under tensile and compressive
strain shifting in opposite directions. Strain is thus widely used
and applied in nanosystems and here we put forward that it
intrinsically leads to spin-orbit effects. We apply this
theoretical framework to both curved two-dimensional systems and
one-dimensional curved quantum wires, showing that in the latter
case two different Rashba-like spin-orbit interactions come into
play enhancing spin-orbit splittings.

\section{Curved two-dimensional systems}

In the thin-wall quantization procedure, quantum excitation
energies in the normal direction are raised far beyond those in
the tangential direction. This allows to neglect the quantum
motion in the normal direction and derive an effective,
dimensionally reduced, Schr\"odinger equation. As opposed to a
classical particle, a quantum particle constrained to a curved
surface retains some knowledge of the surrounding 3D space. In
spite of the absence of interactions, it indeed experiences the
well-known attractive QGP \cite{dac82}. It has been shown that the
JKC thin-wall quantization procedure to derive the effective
Schr\"odinger equation is well-founded, also in the presence of
externally applied electric and magnetic fields\cite{fer08,ort11}.
Empirical evidence for the validity of this approach is provided
by the experimental realization of an optical analogue of the
curvature-induced geometric potential\cite{sza10}.

\begin{figure}[t]
\begin{center}
\includegraphics[width=0.43\textwidth]{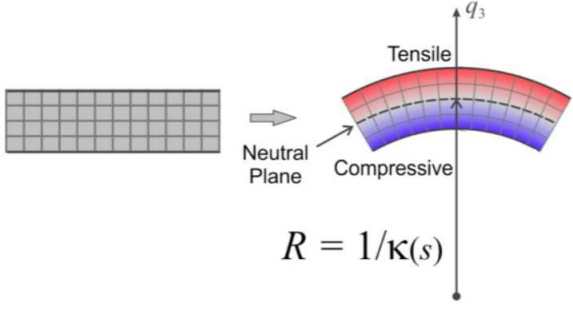}
\end{center}
\caption{Cross
section of a portion of the bent nanostructure with negative value
of the radius of curvature. The effect of curvature leads to
regions under tensile strain for $q_3>0$ and compressive one for
$q_3<0$ }\label{fig:fig1}
\end{figure}

In order to derive the effective Hamiltonian for a curved
two-dimensional electron gas with spin-orbit interaction (SOI), we
therefore use the same conceptual framework of the JKC thin-wall
quantization procedure \cite{jen71}. The mathematical description
is set by defining a 3D curvilinear coordinate system for the
tubular nanostructure. We parameterize the stress-free surface
${\cal S}$ as ${\bf r}={\bf r}(s,z)$ where $z$ indicates the
coordinate along the translationally invariant direction of the
the tubular nanostructure and $s$ corresponds to the arclength
along the curved direction of the thin film. The entire portion of
the nanostructure can be similarly parameterized as ${\bf
R}(s,z,q_3)={\bf r}(s,z) + q_3 \hat{N}(s)$ where $\hat{N}(s)$
indicates the unit vector normal to the bent direction of the thin
film and is oriented so that regions with $q_3 > 0$ are under
compressive strain for positive values of the radius of curvature
and under tensile strain for negative ones [see
Fig.\ref{fig:fig1}]. The strain distribution of the thin film can
be easily derived assuming the in-plane strain
condition\cite{cho92} ${\epsilon}_{zz} \equiv 0$. This can be
justified in nanostructures with a characteristic dimension along
the $z$  direction much larger than the structural dimensions in
the remaining coordinates. Following Landau \cite{lan86}, the
strain component along the bent structure of the thin film reads
$\epsilon_{s\,s} = -q_3 \kappa(s)$ where $\kappa(s)$ indicates the
only non-vanishing principal curvature of the curved stress-free
two-dimensional manifold. The remaining strain component in the
normal direction $q_3$  is related to $\epsilon_{s s}$ by
$\epsilon_{q_{3} q_{3} } = - \left [ \nu / ( 1 - \nu) \right]
\epsilon_{s s}$ with $\nu$ the Poisson ratio.

As the linear deformation potential theory \cite{van89} predicts a
strain-induced shift of the conduction band, local variations of
the strain thus render a local potential for the conducting
electrons ${\cal V}_{\epsilon}(s, q_3) = \gamma \kappa(s) q_3$
which always implies an attraction towards the tensile regions of
the thin film. Here we have introduced the characteristic energy
scale $\gamma$, proportional to the hydrostatic deformation
potential of the conduction valley of the nanostructure, which
typically lies  in the eV scale for conventional semiconductors
\cite{van89}. As the strain field produces an asymmetric confining
potential along the normal direction of the bent surface, it will
immediately yield an average electric field $E \, \hat{N}(s)$
whose strength is proportional to the curvature of the
nanostructure. The strain field therefore leads to a
curvature-induced Rashba SOI. The resulting Schr\"odinger equation
in the effective mass approximation will thus read
\begin{equation}
-\dfrac{\hbar^2}{2 m^{\star}}  G^{i j} {\cal D}_i {\cal D}_j \psi
+ {\cal V}_{\epsilon}(s, q_3) \psi + \dfrac{\alpha}{\hbar} \,
{\boldsymbol \sigma} \cdot {\bf p} \, \times {\hat N(s)} \psi   =
E \psi. \label{eq:eq1}
\end{equation}
where we adopted Einstein summation convention, ${\boldsymbol
\sigma}$ is the Pauli matrix vector and ${\bf p}$ the ordinary
momentum operator in Cartesian coordinates. In Eq.\ref{eq:eq1},
$G^{i j}$ corresponds to the 3D metric tensor which for our
coordinate system takes the simple diagonal form
\begin{equation}
G_{i j} = \left( \begin{array}{ccc}
H(s)^2  & 0 & 0 \\
0 & 1 & 0 \\
0 & 0 & 1 \end{array}
\right) ,
\end{equation}
with $H(s) = 1 - \kappa(s) q_3$. Finally the covariant derivative
${\cal D}_i$ is defined as ${\cal D}_i = \partial_i v_j -
\Gamma_{i \, j }^k v_k$ with $v_j$ the covariant components of a
generic 3D vector field, and $\Gamma_{i \, j }^k$ representing the
Christoffel symbols $$\Gamma_{i j}^{k} = \dfrac{1}{2} G^{k l}
\left[ G_{i l , i} + G_{i l , j} - G_{i i , j} \right] .$$ With
this, the Schr\"odinger equation Eq.~\ref{eq:eq1} can be simply
expanded in our curvilinear coordinate system as
\begin{eqnarray}
E \, \psi&=&\left[-\dfrac{\hbar^2}{2 m^{\star} \, H(s)}  \partial_s \left(\dfrac{1}{H(s)} \partial_s\right)-\dfrac{\hbar^2}{2 m^{\star} \, H(s)} \partial_{q_{3}} \right. \nonumber \\ & & \left.  \left(H(s) \, \partial_{q_{3}}\right) -\dfrac{\hbar^2}{2 m^{\star}} \partial^2_y+ {\cal V}_{\epsilon}(s,q_3)+ {\cal V}_\lambda(q_3) \right] \psi \nonumber \\ & &  - i \alpha \left( \sigma_s \partial_y - \dfrac{\sigma_y}{H(s)} \partial_s \right) \psi,
\end{eqnarray}
where $\sigma_{s} = {\boldsymbol \sigma} \cdot {\hat{s}}$ and the
momentum along the bent direction of the nanostructure is $p_s
\equiv - i \hbar  G^{s \, s}  \partial_s$ .  We have also
introduced a squeezing potential in the normal direction ${\cal
V}_\lambda (q_3)$. In Ref.~\cite{ort11b} this was assumed to be
given by two infinite step potential barriers at $\pm \delta / 2$
with $\delta$ the total thickness of the thin film. In the
remainder we will instead consider the case of an harmonic trap
${\cal V}_{\lambda} (q_3) = \lambda q_3^2 / 2$ to show that the
ensuing results are not dependent on the specific form of the
squeezing potential at hand. To proceed further we introduce, in
the same spirit of the thin-wall quantization procedure, a
rescaled wavefunction $\chi = \psi \times \sqrt{H(s)}$ for which
the surface density probability is defined as  $\int |
\chi(s,y,q_3) |^2 d q_3$. The resulting Schr\"odinger equation is
then determined by an effective Hamiltonian
\begin{eqnarray}
{\cal H}&=&-\dfrac{\hbar^2}{2 m^{\star}} \partial_z^2  - \dfrac{\hbar^2}{2 m^{\star}} \dfrac{\partial_s^2 }{H(s)^2}+\dfrac{\hbar^2}{m^{\star}} \dfrac{\partial_s H(s) \, \partial_s}{H(s)^3} \nonumber \\ & & -\dfrac{\hbar^2}{2 m^{\star}} \partial_{q_{3}}^2
-\dfrac{\hbar^2}{2 m^{\star}} \, \left[\dfrac{5}{4} \, \dfrac{\left(\partial_s H(s) \right)^2}{H(s)^4}-\dfrac{\partial_s^2 H(s)}{2 \, H(s)^3} \right] \nonumber \\  & &-\dfrac{\hbar^2}{2 m^{\star}} \dfrac{\left(\partial_{q_{3}} H(s) \right)^2}{4 \, H(s)^2} + {\cal V}_{\lambda} (q_3) + {\cal V}_{\epsilon}(s,q_3)  \nonumber \\ & & - i \alpha \left(\sigma_s \partial_z - \sigma_z \dfrac{\partial_s}{H(s)} \right) + i \alpha \sigma_z  \dfrac{\partial_s H(s)}{H(s)^{\frac{3}{2}}} . \label{eq:effectivehamiltonian}
\end{eqnarray}
For large enough strength of the squeezing potential in the normal
direction, we can follow Ref.~\cite{ort11b} and expand the
Hamiltonian of Eq.\ref{eq:effectivehamiltonian} as ${\cal
H}=\sum_k q_3^k {\cal H}_{k}$. At the zeroth order in $q_3$ and
retaining the leading order correction linear in $q_3$, the
effective Hamiltonian reads:
\begin{eqnarray} {\cal H}_0 &=&
-\dfrac{\hbar^2}{2 m^{\star}} \partial_z^2  - \dfrac{\hbar^2}{2
m^{\star}} \partial_s^2 - \dfrac{\hbar^2}{2 m^{\star}}
\partial_{q_3}^2 -\dfrac{\hbar^2 \kappa(s)^2}{8 m^{\star}}
\nonumber \\ & & \nonumber \\ & &  - i \alpha \left(\sigma_s
\partial_z - \sigma_z \partial_s \right)+ {\cal V}_{\lambda}(q_3)
+ {\cal V}_{\epsilon}(s,q_3). \label{eq:hamiltonian0}
\end{eqnarray}
The strong size quantisation along the normal direction allows to
employ the adiabatic approximation and consider an ansatz for the
wave function $\chi(s, z, q_3) = \chi^T (s, q_z) \times \chi^N (s,
q_3)$ with $\chi_N(s,q_3)$ solving at fixed $s$ the
one-dimensional Schr\"odinger equation for the fast normal degrees
of freedom which is regulated by the Hamiltonian
\begin{equation}
{\cal H}_0^N = - \dfrac{\hbar^2}{2 m^{\star}} \partial_{q_3}^2 +
\gamma \kappa(s) q_3 + \dfrac{1}{2} \lambda q_3^2.
\end{equation}
The spectrum of the Hamiltonian above can be easily read as $E^N = \hbar \sqrt{\lambda / m} \left( n + 1/2 \right) - \gamma^2 \kappa(s)^2 / \lambda$.
With this, the effective Hamiltonian for the slow tangential quantum degrees of freedom can be found by integrating out the $q_3$ degrees of freedom.
We then get the dimensionally reduced effective Hamiltonian
\begin{equation}
{\cal H}_0^T =  -\dfrac{\hbar^2}{2 m^{\star}} \partial_z^2  - \dfrac{\hbar^2}{2 m^{\star}} \partial_s^2  -\dfrac{S^2 \kappa(s)^2}{8 m^{\star}}
- i \alpha \left(\sigma_s \partial_z - \sigma_z \partial_s \right).
\label{eq:hamiltoniantangential}
\end{equation}
It has the same functional form of the effective Hamiltonian for a
planar two-dimensional electron gas with Rashba spin-orbit
interaction and a geometric potential whose magnitude $S^2 =
\hbar^2 + 8 m^{\star} \gamma^2 /  \lambda$ is strongly
renormalised by the strain-induced geometric potential
\cite{ort11b}. We point out that for systems with an intrinsic SOI
the normal and tangential degrees of freedom, curvature induced
geometric potential proportional to the mean curvature of the
manifold can appear ~\cite{chang2013}. These terms are absent in
the strain-induced Rashba spin-orbit coupling because of the
presence of the normal potential gradient.

We next use this theoretical framework to derive the
effective dimensionally reduced Hamiltonian considering  the case
of a rolled-up cylindrical nanotube and a nanocorrugated thin
film. For a cylindrical nanotube of radius $R$  the arclength can
be easily expressed in cylindrical coordinates as $s = R \phi$ and
the resulting effective Hamiltonian for a cylindrical
two-dimensional electron gas (C2DEG) then reads \cite{tru07}
\begin{eqnarray}
{\cal H}_{C2DEG}& = &  -\dfrac{\hbar^2}{2 m^{\star}} \partial_z^2  - \dfrac{\hbar^2}{2 m^{\star} R^2} \partial_\phi^2 \nonumber \\ & & - i \alpha \left(\sigma_{\phi} \partial_z - \sigma_z \dfrac{\partial_{\phi}}{R}  \right),
\end{eqnarray}
where we have neglected the geometric potential since it
corresponds to a rigid shift of the energies. For a nanocorrugated
thin film instead, we start out by parametrizing the stress-free
surface in the Monge gauge as $y=h(x)$ with $h$ indicating the
height of the corrugation with respect to its planar counterpart.
In Cartesian coordinates then the effective Hamiltonian
reads
\begin{eqnarray} {\cal H}_{NC}& = & -\dfrac{\hbar^2}{2
m^{\star}} \partial_z^2  - \dfrac{\hbar^2}{2 m^{\star}} \dfrac{1}{
1 + h^{\prime}(x)^2}  \partial_x^2 \nonumber \\ & & -
\dfrac{S^2}{8 m^{\star}} \dfrac{ h^{\prime \prime}(x)^2}{\left[1 +
h^{\prime}(x)\right]^3}  \\ & & -  \dfrac{i \alpha}{\sqrt{1 +
h^{\prime}(x)^2}}  \left(\sigma_x  \partial_z + \sigma_y
h^{\prime}(x) \partial_z - \sigma_z \partial_x \right) \nonumber
\end{eqnarray}

\section{Curved quantum wires}

We now apply the  theoretical framework we developed in  the
previous section to planar curved one-dimensional electron gases.
In this case, due to the asymmetric confinement in the direction
perpendicular to the plane of the curved one-dimensional systems
($\hat z$ following the definition of the curvilinear coordinate
system given in Sec.II) there is an additional Rashba SOI
different from the strain-induced one. The effective Schr\"odinger
equation reads
\begin{eqnarray} E \psi &=& -\dfrac{\hbar^2}{2 m^{\star}}
G^{i j} {\cal D}_i {\cal D}_j \psi + {\cal V}_{\epsilon}(s, q_3)
\psi + \nonumber \\ & & \dfrac{\alpha}{\hbar} \,  {\boldsymbol
\sigma} \cdot {\bf p} \, \times {\hat N(s)} \psi   + (
\dfrac{\alpha_R}{\hbar} \,  {\boldsymbol \sigma} \cdot {\bf p} \,
\times {\hat z} ) \psi \label{eq:eq2}
\end{eqnarray}
Since there is no coupling among the curvature of the one-dimensional planar wire and the direction normal to the plane, we can safely neglect the degrees of freedom in its direction.
By expanding Eq.\ref{eq:eq2} by covariant calculus we get the Schr\"odinger equation in the planar two-dimensional space as
\begin{eqnarray}
E \, \psi&=&\left[-\dfrac{\hbar^2}{2 m^{\star} \, H(s)}  \partial_s \left(\dfrac{1}{H(s)} \partial_s\right)-\dfrac{\hbar^2}{2 m^{\star} \, H(s)} \partial_{q_{3}} \right. \nonumber \\ & & \left. \left(H(s) \, \partial_{q_{3}}\right) + {\cal V}_{\epsilon}(s,q_3)+ {\cal V}_\lambda(q_3) \right] \psi  \\ & & + i \alpha \left(  \dfrac{\sigma_z}{H(s)} \partial_s \right) \psi - i \alpha_R  \left(  \dfrac{\sigma_N}{H(s)} \partial_s  - \sigma_s \partial_{q_3} \right) \psi \nonumber ,
\end{eqnarray}
where $\sigma_N = {\boldsymbol \sigma} \cdot \hat{N}(s)$.

\begin{figure}[t]
\begin{center}
\includegraphics[width=0.4\textwidth]{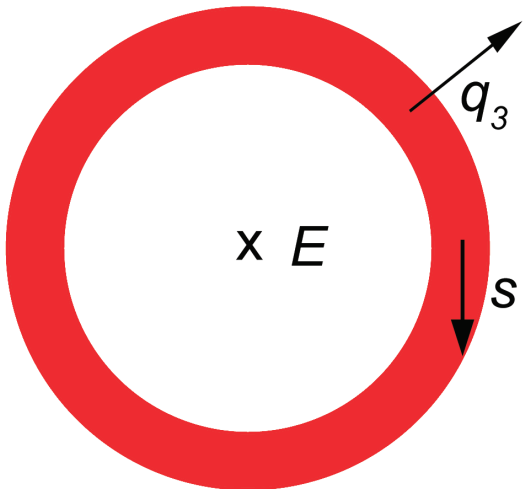}
\end{center}
\caption{Schematics of the ring nanostructure. The electric field
in the direction normal to the plane is due to an asymmetric
confinement. Strain effects instead render an  electric field in
the direction $q_3$ normal to the ring.}\label{fig:fig3}
\end{figure}

The effective dimensionally reduced one-dimensional Schr\"odinger
equation can be found by first performing a scaling of the
wavefunction so that the one-dimensional surface density
probability reads $\int  | \chi(s,q_3)|^2  d q_3$, then expanding
the resulting Hamiltonian as in Sec.II and finally employing the
method of adiabatic separation among the fast normal quantum
degree of freedom $q_3$ an the planar slow quantum degree of
freedom $s$. The resulting effective Hamiltonian is
\begin{eqnarray}
{\cal H}_0 &=& - \dfrac{\hbar^2}{2 m^{\star}} \partial_s^2 -  \dfrac{S ^2 \kappa(s)^2}{8 m^{\star}} + i \alpha \sigma_z \partial_s  \nonumber \\ & &  - i \alpha_R \left(\sigma_N   \partial_s  - \dfrac{1}{2} \sigma_s \kappa(s)  \right),
\label{eq:hamiltonian01D}
\end{eqnarray}
which agrees with the Hamiltonian for a curved one-dimensional
quantum wire proposed in Ref.~\cite{zha07} once the strain-induced
Rashba SOI is neglected. To analyse the influence of the latter,
we now consider the example of a closed quantum ring \cite{mei02}
shown schematically in Fig.\ref{fig:fig3} .
%
%
\begin{figure}[t]
\begin{center}
\includegraphics[width=0.4\textwidth]{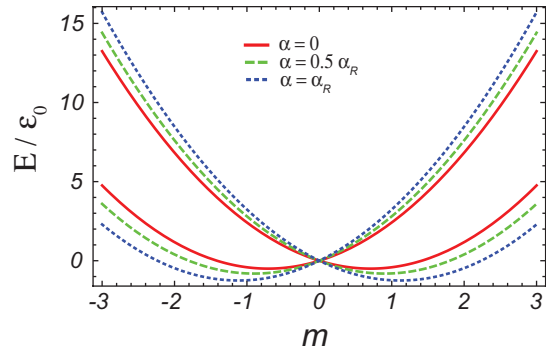}
\end{center}
\caption{The bandstructure for a closed ring with strength of the
Rashba SOI in the direction perpendicular to the plane $\alpha_R =
\hbar^2 / (2 m^{\star} R)$ and different values of the
strain-induced Rashba SOI $\alpha$. Energies are measured in units
of $\epsilon_0 = \hbar^2 / ( 2 m^{\star} R^2)$. The spin-orbit
splitting is generally enhanced by the effect of the Rashba SOI.}
\label{fig:fig4}
\end{figure}
Adopting polar coordinates we have
\begin{eqnarray}
\sigma_s &=& \sigma_x \sin{\phi} - \sigma_y \cos{\phi} \nonumber \\
\sigma_N &=&   \sigma_x \cos{\phi} + \sigma_y \sin{\phi},
\end{eqnarray}
whereas $s = R \phi$. With this, the effective Hamiltonian for the
1D ring can be then solved using a trial spinorial wave function
of the form $ \chi = \left[ \chi_1 \mathrm{e}^{i (m - 1 / 2) \phi}
, \chi_2 \mathrm{e}^{i (m + 1 / 2 ) \phi } \right]^{T}$ where $m$
can only assume half-integer values to fulfill the periodic
boundary conditions. The amplitude $\chi_1$ and $\chi_2$ depends
on $m$. The corresponding energy spectrum can be simply found as
\begin{eqnarray}
E(m)& =&  \dfrac{\hbar^2}{2 m^{\star} R^2} (m^2+ \dfrac{1}{4}) + \dfrac{\alpha}{2 R} \nonumber \\ & &   \pm |m| \sqrt{\left( \dfrac{\hbar^2}{2 m^{\star} R^2}  + \dfrac{\alpha}{R} \right)^2 + \dfrac{\alpha_R^2}{R^2}} .
\end{eqnarray}
As shown in Fig.\ref{fig:fig4} the strain-induced Rashba SOI
enhances the spin-orbit splitting due to the Rashba SOI in the
direction perpendicular to the plane. The spin properties of the
eigenstates of the Hamiltonian in Eq. \ref{eq:hamiltonian01D} are
not modified by the presence of the strain-induced Rashba SOI.
Indeed, the expectation values for $\sigma_s$ and $\sigma_N$ are
$\langle \chi| \sigma_s |\chi \rangle = 0$ and $\langle \chi |
\sigma_N |\chi \rangle = 2 \chi_1 \, \chi_2 $, respectively. This
implies that the averaged spin projection, as expected, is
pointing perpendicularly to the ring along the radial direction
while the tangential component is vanishing. There is also a non
trivial $z-$projection which is given by $\langle \chi| \sigma_z
|\chi \rangle = \chi_1^2- \chi_2^2 $ which is a characteristic
signature of curvature effects. Hence, the presence of the
strain-induced Rashba SOI affects only the intensity of the
averaged spin projections by modifying the spinorial components of
the wave function $\chi$. In particular, such term tends to
enhance the out-of-plan spin projection. We expect this
contribution to amplify the curvature effects on the spin
transport properties of quantum rings~\cite{liu}.

\section{Conclusions}
We have derived, in conclusion, a dimensionally reduced
Schr\"odinger equation with spin-orbit interaction in
two-dimensional and effective one-dimensional electronic
strain-driven nanostructures. By employing a method of adiabatic
separation of fast and slow quantum degrees of freedom, we have
shown that the effects of a finite curvature are twofold. First,
in agreement with Ref.~\cite{ort11b}, the strain effects render an
often gigantic renormalisation of the curved-induced quantum
geometric potential. Second, the asymmetric confinement due to the
strain field leads to a Rashba-type spin-orbit interaction whose
strength is proportional to the local curvature of the
nanostructure. Applying this theoretical framework to
one-dimensional curved quantum wires leads to an enhanced
spin-orbit splitting due to the presence of two Rashba-type SOI.
The inclusion of strain effects therefore boosts the SOI which
will strongly affect the electron spin transport properties of
strain-driven nano structures in curved geometries.

\section*{Acknowledgments} \noindent We thank Jeroen van den Brink
for fruitful and stimulating discussions. The research leading to
these results has received funding from the FP7/2007-2013 under
grant agreement N. 264098 - MAMA.

\end{document}